\documentclass[aps, pra, twocolumn, amsmath, amssymb, superscriptaddress, nofootinbib]{revtex4-1}

\makeatletter
\def\p@subsection{}
\def\p@subsubsection{}
\makeatother

\usepackage[utf8]{inputenc}
\usepackage{graphicx}
\usepackage{units}
\usepackage{color}
\usepackage[pdftex, colorlinks=true, linkcolor=myblue, citecolor=myblue,
urlcolor=myblue]{hyperref}
\usepackage{times}
\usepackage{comment}
\usepackage{graphicx}
\usepackage{amsmath, calc}
\usepackage{mathrsfs}
\usepackage{amsfonts}
\usepackage{amssymb}
\usepackage{bm}
\usepackage{bbm}
\usepackage{color}
\usepackage{array}
\usepackage{units}
\usepackage{tabstackengine}
\stackMath

\definecolor{myblue}{rgb}{0,0,1}
\definecolor{myred}{rgb}{1,0,0}

\DeclareMathOperator{\Tr}{Tr}

\newcommand\scalemath[2]{\scalebox{#1}{\mbox{\ensuremath{\displaystyle #2}}}}

\begin{document}
%\narrowtext

%===========================================================================
%===========================================================================
%===========================================================================
%===========================================================================
%===========================================================================
%===========================================================================
%===========================================================================
%===========================================================================

\title{A quantum battery with quadratic driving}

%===========================================================================
%===========================================================================
%===========================================================================
%===========================================================================
%===========================================================================
%===========================================================================
%===========================================================================
%===========================================================================

\author{Charles Andrew Downing*} 
\affiliation{Department of Physics and Astronomy, University of Exeter, Exeter EX4 4QL, United Kingdom}

\author{Muhammad Shoufie Ukhtary}
\affiliation{Department of Physics and Astronomy, University of Exeter, Exeter EX4 4QL, United Kingdom}
\affiliation{Research Center for Quantum Physics, National Research and Innovation Agency (BRIN), South Tangerang 15314, Indonesia
\\ *c.a.downing@exeter.ac.uk}

%===========================================================================
%===========================================================================
%===========================================================================
%===========================================================================
%===========================================================================
%===========================================================================
%===========================================================================
%===========================================================================

\date{\today}

%===========================================================================
%===========================================================================
%===========================================================================
%===========================================================================
%===========================================================================
%===========================================================================
%===========================================================================
%===========================================================================

\begin{abstract}
\noindent \textbf{Abstract}\\
Quantum batteries are energy storage devices built using quantum mechanical objects, which are developed with the aim of outperforming their classical counterparts. Proposing optimal designs of quantum batteries which are able to exploit
quantum advantages requires balancing the competing demands for fast charging, durable storage and effective work extraction. Here we study theoretically a bipartite quantum battery model, composed of a driven charger connected to an energy holder, within two paradigmatic cases of a driven-dissipative open quantum system: linear driving and quadratic driving. The linear battery is governed by a single exceptional point which splits the response of the battery into two regimes, one of which induces a good amount of useful work. Quadratic driving leads to a squeezed quantum battery, which generates plentiful useful work near to critical points associated with dissipative phase transitions. Our theoretical results may be realized with parametric cavities or nonlinear circuits, potentially leading to the manifestation of a quantum battery exhibiting squeezing.
\end{abstract}

%===========================================================================
%===========================================================================
%===========================================================================
%===========================================================================
%===========================================================================
%===========================================================================
%===========================================================================
%===========================================================================

\maketitle

%\pacs{73.22.Pr, 73.21.La, 03.65.Ge, 03.65.Pm}

%===========================================================================
%===========================================================================
%===========================================================================
%===========================================================================
%===========================================================================
%===========================================================================
%===========================================================================
%===========================================================================

\noindent \textbf{Introduction}\\
The Laws of Thermodynamics allow for a complete description of classical thermal machines, from classical heat engines to refrigerators. However, the ongoing trend for device miniaturization inevitably led to quantum effects becoming important. This realization required the development of theories of thermal energy conversion in the quantum regime~\cite{Kosloff2014, Bhattacharjee2021}. Shortly afterwards, an influential study of energy storage and transduction at the nanoscale pioneered the concept of a quantum system storing and releasing energy on demand: a quantum battery~\cite{Alicki2013, Campaioli2023}.

An archetypal quantum battery model consists of two parts, the battery holder and the battery charger. The holder is essentially isolated from the external environment in order to prevent energy loss, and hence it is modelled as a dissipationless subsystem. In order to receive energy, the battery holder then needs to be coupled to the the battery charger. The charger subsystem is supposed to feel the environment -- allowing it to be driven -- but this comes at the cost of the charger suffering from dissipation. After charging for some finite period of time the battery holder is disconnected from the battery charger so that energy storage, and eventually energy extraction on demand, may occur~\cite{Farina2019}. This fundamental bipartite quantum battery model, along with its charging, storage and discharging performances, has been considered theoretically in various guises over the last few years of intensive quantum battery research~\cite{Ferraro2018, Andolina2018, Le2018, Barra2019, Zhang2019, Santos2019, Pirmoradian2019, Keck2019, Crescente2020, Santos2020, Carrega2020, Santos2021, Xu2021, Dou2022, Barra2022, Carrasco2022, Shi2022, Santos2023}.

Experimentally, superabsorption in a  Dicke-style quantum battery has already been reported~\cite{Quach2022}, as has a detailed investigation of collective charging in a variety of spin-based systems~\cite{Joshi2022}. The realization of a quantum battery based upon a superconducting qutrit, including a characterization of its charging and self-discharging processes, has also been achieved in some recent and rather ingenious experiments~\cite{Hu2022}. Meanwhile, the first steps in gauging the performance of quantum emitters and light fields for the purpose of energy transfer have been performed empirically~\cite{Maillette2022}. Complementary studies of the energetics of various quantum objects, including qubits~\cite{Cimini2020, Stevens2022} and nuclear spins~\cite{Peterson2019}, demonstrate the  promise of this nascent quantum technological field.

In what follows, we consider theoretically the celebrated bipartite quantum battery model, where the battery charger and battery holder are modelled as two coupled quantum harmonic oscillators, in two different circumstances. Firstly, we revisit the linear (one-photon) driving setup as discussed by Farina and co-workers~\cite{Farina2019}. This study both lays the theoretical framework for the Gaussian quantum batteries considered and acts as a comparator for the second case of interest: a quadratic (two-photon) driving arrangement~\cite{Leghtas2015, Wang2016, Pechal2019, Wustmann2019, Gaikwad2023}. The latter case of parametric driving significantly alters the underlying physics of the quantum battery due to the inducement of spectral collapse, quantum squeezing, dynamic instabilities and dissipative phase transitions. Most importantly, we reveal that the quadratic quantum battery allows for abundant useful energy to be stored in it when the driving amplitude is near to certain critical values.
\\

\begin{figure*}[tb]
\begin{center}
 \includegraphics[width=1.0\linewidth]{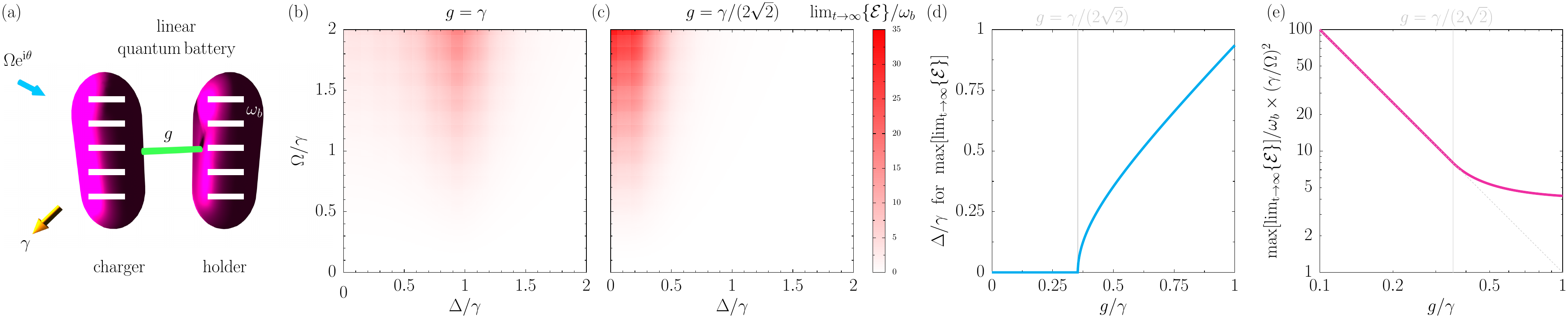}
 \caption{ \textbf{The linear quantum battery and its steady state.} Panel (a): a sketch of the bipartite quantum battery (both parts are modelled as quantum harmonic oscillators, with the level spacing $\omega_b$), composed of a battery charger coupled to a battery holder with the coupling strength $g$ (green bar). The charger is driven coherently (cyan arrow) with an amplitude $\Omega$, phase $\theta$ and frequency $\omega_d$, while it suffers loss (yellow arrow) at the decay rate $\gamma$. Panel (b): the ergotropy $\mathcal{E}$ in the steady state (in units of $\omega_b$) as a function of $\Omega$ and $\Delta$ (both in units of $\gamma$) from Eq.~\eqref{eq:sfdsdfwwwq}. Here the coupling $g = \gamma$. Panel (c): as for panel (b), but with $g = \gamma/(2\sqrt{2}) \simeq 0.35 \gamma$. Panel (d): the detuning $\Delta$ associated with the maximum of the steady state ergotropy $\mathcal{E}$, as a function of the coupling $g$ [cf. Eq.~\eqref{eq:dfdfdfdfaa}]. Panel (e): a semi-logarithmic plot of the maximum of the steady state ergotropy $\mathcal{E}$, in units of $\omega_b$ and scaled by $(\gamma/\Omega)^2$, as a function of the coupling $g$ [cf. Eq.~\eqref{eq:dfddfffdfdfaa}].}
 \label{Figure1}
 \end{center}
\end{figure*}

\noindent \textbf{Results}\\
\noindent \textit{Bipartite quantum battery model}\\
The total Hamiltonian operator $\hat{H}$ of the composite quantum battery system can be decomposed into four parts,
\begin{equation}
\label{eq:Haxcsdfdsfdsfsdfvcxvmy}
 \hat{H} =  \hat{H}_c +  \hat{H}_h +  \hat{H}_{c-h} +  \hat{H}_{d},
\end{equation}
which accounts for the battery charger energy, the battery holder energy, the charger-holder coupling and the coherent driving of the charger respectively. Taking $\hbar = 1$ throughout, these Hamiltonian contributions are -- each in turn -- defined by
    \begin{align}
    \label{eq:sfsfagttt}
 \hat{H}_c  &= \omega_b \, c^\dagger c, \\
 \hat{H}_h  &= \omega_b \, h^\dagger h, \\
\hat{H}_{c-h} &= g \, \left( c^\dagger h + h^\dagger c \right),     \label{eq:sfsdfgsdghbnnfagttt} \\
\hat{H}_{d} &= \Omega  \mathrm{e}^{\mathrm{i} \theta} \mathrm{e}^{-\mathrm{i} \omega_d t} \, c^\dagger + \mathrm{h. c.\, }. \label{eq:lksdsad}
 \end{align}
The energy level spacings of the harmonic oscillators modelling the battery charger and the battery holder are equal at $\omega_b$, the coupling strength between them is $g$, and the laser driving the charger has an amplitude $\Omega \ge 0$, phase $\theta$ and frequency $\omega_d$. Excitations in the battery charger are created and destroyed by the operators $c^\dagger$ and $c$ respectively, while the ladder operators $h^\dagger$ and $h$ track the transitions up and down the energy ladder of the battery holder. Both flavours of operator are subject to bosonic commutation relations, $[c, c^\dagger] = 1$ and $[h, h^\dagger] = 1$. Notably, the counter-rotating terms $\propto c h$ and $\propto c^\dagger h^\dagger$ not appearing in the coupling Hamiltonian of Eq.~\eqref{eq:sfsdfgsdghbnnfagttt} have been dropped since they are small for typical couplings $g$ satisfying $g \ll \omega_b$. Therefore, we do not enter the so-called ultrastrong coupling regime, as commonly defined without reference to losses~\cite{Diaz2019, Boite2020, Toghill2022, Shaghaghi2002} (see also Supplementary Notes 2 and 3, where we find that this rotating-wave approximation is a good one for any coupling $g$ satisfying $g \lesssim \omega_b/1000$). A graphical representation of this bipartite battery arrangement, as captured mathematically by Eq.~\eqref{eq:Haxcsdfdsfdsfsdfvcxvmy}, is sketched in Fig.~\ref{Figure1}~(a). This cartoon includes the three key parameters of the model: the driving amplitude $\Omega$, the charger-holder coupling $g$, and the loss rate of the charger $\gamma$.

While the battery holder is approximated as a dissipationless subsystem, we consider the battery charger subsystem to suffer loss (as measured with the decay rate $\gamma$, where $\gamma \ge 0$).  Within an open quantum systems approach~\cite{Breuer2002, Downing2022}, we employ the Gorini–Kossakowski–Sudarshan–Lindblad quantum master equation for the density matrix $\rho$, in the standard form $\partial_t \rho = \mathrm{i} [ \rho,  \hat{H} ] + (\gamma/2 ) \mathcal{L}_c \left[ \rho \right]$. Here the Lindbladian superoperator $\mathcal{L}_c \left[ \rho \right] = 2 c \rho c^\dagger - c^\dagger c \rho - \rho c^\dagger c$ acts on $\rho$, while the unitary dynamics of the closed system are provided by the Hamiltonian operator $\hat{H}$ of Eq.~\eqref{eq:Haxcsdfdsfdsfsdfvcxvmy}. In arriving at this master equation, we have employed the Born approximation (due to the assumption of weak interactions between the system and the environment), the Markov approximation (due to the supposition that the memory of the environment is much shorter than that of the system), and the secular approximation (due to certain transition frequencies leading to quickly-rotating terms which are neglectable)~\cite{Breuer2002}.

 In order to judge the energetic performance of the quantum battery there are two crucial energies of interest, $E_h$ and $ E_h^{\beta}$, which are both associated with the battery holder~\cite{Farina2019}
     \begin{align}
         \label{eq:sfssdfsdfsdfagttt}
  E_h  &=  \Tr \{ \hat{H}_h \rho_h \}, \\
  E_h^{\beta} &= \Tr \{ \hat{H}_h \rho_h^{\beta} \},          \label{eq:sfdfsdujj}
 \end{align}
 here the reduced density matrix $\rho_h$ of the battery holder subsystem $h$ is defined by $\rho_h = \Tr_c \{ \rho \}$, where the partial trace over the battery charger subsystem $c$ has been taken. The quantity $E_h$ in Eq.~\eqref{eq:sfssdfsdfsdfagttt} measures the mean energy stored in the battery holder due to the state $\rho_h$. However, not all of this energy $E_h$ may be useful (for example, the system may be in thermal equilibrium). The state $\rho_h^{\beta}$ appearing in Eq.~\eqref{eq:sfdfsdujj} is the so-called passive state of the battery holder, which has the property that no work can be extracted from it cyclically under unitary evolution~\cite{Pusz1977, Lenard1978, Friis2018}. The passive state $\rho_h^{\beta}$ can be obtained by re-ordering the eigenvalues of the Hamiltonian and density matrix appropriately~\cite{Allahverdyan2004}. The energies of Eq.~\eqref{eq:sfssdfsdfsdfagttt} and Eq.~\eqref{eq:sfdfsdujj} may then be simply combined into an influential measure -- the so-called ergotropy $\mathcal{E}$ -- like so
     \begin{equation}
              \label{eq:sfdhaaaarr}
    \mathcal{E}  = E_h - E_h^{\beta}, 
 \end{equation}
 which measures the upper bound of the useful energy stored within the battery holder~\cite{Allahverdyan2004}. The ergotropy $\mathcal{E}$ will be nonzero if the state $\rho_h$ is non-passive, which can arise due to population inversion or thanks to certain coherences for example. Understanding the dynamical and steady state behaviors of the ergotropy $\mathcal{E}$ for the bipartite quantum battery, with linear and quadratic drivings respectively, is the principle goal of this theoretical study.

For the purposes of the following calculations, it is convenient to work with the rotated density matrix $\rho \to \tilde{\rho}$, found via the transformation $\tilde{\rho} = \mathrm{e}^{\mathrm{i} \omega_d t \left( c^\dagger c + h^\dagger h \right) } \rho \mathrm{e}^{-\mathrm{i} \omega_d t \left( c^\dagger c + h^\dagger h \right)}$. This leads to the quantum master equation for $\tilde{\rho}$, complete with the effective Hamiltonian operator $\hat{\mathcal{H}}$, as follows [cf. Eqs.~\eqref{eq:sfsfagttt}--\eqref{eq:lksdsad}]
    \begin{align}
          \label{eq:mnnbfgvd}
\partial_t \tilde{\rho} =&~\mathrm{i} \left[ \tilde{\rho}, \hat{\mathcal{H}} \right] + \frac{\gamma}{2} \mathcal{L}_c \left[ \tilde{\rho} \right], \\
\hat{\mathcal{H}} =&~\Delta \left( c^{\dagger} c + h^{\dagger} h \right) + g \left( c^{\dagger} h + h^{\dagger} c \right) \nonumber \\
&+ \Omega \left(  \mathrm{e}^{\mathrm{i} \theta} c^\dagger + \mathrm{e}^{-\mathrm{i} \theta} c \right),      \label{eq:mnnbfgvd2}
 \end{align}
where we have introduced the driving-battery detuning frequency $\Delta = \omega_b - \omega_d$. The rotation of the reduced density matrix $\rho_h \to \tilde{\rho}_h$ is similarly governed by $\tilde{\rho}_h = \mathrm{e}^{ \mathrm{i} \omega_d t \,  h^\dagger h  } \rho_h \mathrm{e}^{- \mathrm{i} \omega_d t \,  h^\dagger h }$, such that the energies of interest [cf. Eq.~\eqref{eq:sfssdfsdfsdfagttt} and Eq.~\eqref{eq:sfdfsdujj}] can be equivalently written as $E_h  =  \Tr \{ \hat{H}_h \tilde{\rho}_h \}$ and $E_h^{\beta}  =   \Tr \{ \hat{H}_h \tilde{\rho}_h^{\beta} \}$. Furthermore, the properties of traces and partial traces allows for the aforementioned energetic quantities, given in terms of the reduced density matrices $\tilde{\rho}_h$ and $\tilde{\rho}_h^{\beta}$, to be reconfigured in terms of the full (and rotated) density matrix $\tilde{\rho}$ like so: $\Tr \{ \hat{H}_h \tilde{\rho}_h \} = \Tr \{ \hat{H}_h \tilde{\rho} \}$ and $\Tr \{ \hat{H}_h \tilde{\rho}_h^{\beta} \} = \Tr \{ \hat{H}_h \tilde{\rho}^{\beta} \}$. Given this identification, one can use the trace property $\Tr \{ \mathcal{O} \tilde{\rho} \} = \langle \mathcal{O} \rangle$, which is valid for any operator $\mathcal{O}$, to simplify both energies of interest $E_h$ and $E_h^\beta$. We consider the initial state of the system (at $t=0$) to be the product of the local vacuum states of the battery charger and battery holder respectively, $\tilde{\rho} \left( t = 0 \right) = | 0 \rangle \langle 0 |_c  \otimes | 0 \rangle \langle 0 |_h  $, after which both battery components start to interact as the charging process commences.

Pleasingly, one may exploit some neat properties of Gaussian systems~\cite{Ferraro2005, Olivares2012, Serafini2017} in order to fully determine the passive state energy $E_h^{\beta}$ (see Supplementary Note 1 and elsewhere~\cite{Farina2019} for more details). We then finally arrive at compact expressions for both Eq.~\eqref{eq:sfssdfsdfsdfagttt} and Eq.~\eqref{eq:sfdfsdujj}, in terms of operator expectation values, as follows
    \begin{align}
       \label{eq:dpvrdfr}
  E_h  &= \omega_b \langle h^\dagger h \rangle, \\
 E_h^{\beta} &= \omega_b  \left( \tfrac{\sqrt{\mathcal{D}}-1}{2} \right).        \label{eq:dpvrdfr2}
 \end{align}
 Here we have introduced the dimensionless quantity $\mathcal{D}$, which collects the first and second moments of the battery holder, that is objects like $\langle h \rangle$ and $\langle h^\dagger h \rangle$, in the form
     \begin{equation}
            \label{eq:dpvrdfsfdsdfdfr}
\mathcal{D} = \Bigl\{ 1 + 2 \langle h^\dagger h \rangle - 2 \langle h^\dagger  \rangle \langle h \rangle \Bigl\}^2 - 4 \Big| \langle h h  \rangle-\langle h \rangle^2  \Big|^2.
 \end{equation}
Now the ergotropy $\mathcal{E}$ of the battery holder can be readily computed via Eq.~\eqref{eq:sfdhaaaarr}, along with Eq.~\eqref{eq:dpvrdfr}--\eqref{eq:dpvrdfsfdsdfdfr}. In what follows, we consider the dynamical and steady state ergotropies for the composite quantum battery system, firstly with the one-photon driving case of Eq.~\eqref{eq:lksdsad} and then for the arguably more interesting quadratic case, featuring two-photon driving of the battery charger.
\\

%===========================================================================
%===========================================================================
%===========================================================================
%===========================================================================
%===========================================================================
%===========================================================================
%===========================================================================
%===========================================================================

\noindent \textit{Linear quantum battery}\\
The situation with one-photon coherent driving, as sketched in Fig.~\ref{Figure1}~(a) and as governed by the Hamiltonian operator $\hat{\mathcal{H}}$ of Eq.~\eqref{eq:mnnbfgvd2}, readily allows for analysis of the first and second moments (as is carried out in Supplementary Note 2, see also calculations in the literature~\cite{Farina2019} for the more specific case with $\Delta = 0$). Since the system correlators factorize perfectly, such that $\langle h^\dagger h \rangle = \langle h^\dagger \rangle \langle h \rangle$ and $\langle h h \rangle = \langle h \rangle \langle h \rangle$ for example, the key quantity $\mathcal{D}$ as defined in Eq.~\eqref{eq:dpvrdfsfdsdfdfr} simply reduces to
    \begin{equation}
    \label{eq:sfdsdfwrfgsfdwwdvq}
\mathcal{D} = 1.
 \end{equation}
This correlator factorization stems from the joint battery charger--battery holder system evolving in a product state~\cite{Ferraro2005, Olivares2012, Serafini2017}. (Notably, the neat Eq.~\eqref{eq:sfdsdfwrfgsfdwwdvq} is not met when there is nonzero incoherent driving, which leads to unuseful energy storage in that particular case, as discussed previously~\cite{Farina2019}). Given the form of Eq.~\eqref{eq:sfdsdfwrfgsfdwwdvq}, the battery holder state must be pure due to the properties of Gaussian systems~\cite{Serafini2017}. Hence all of the stored energy in the battery holder is useful because the passive state energy must be [cf. Eq.~\eqref{eq:dpvrdfr2}]
     \begin{equation}
    \label{eq:sfdsdfwrfgsfdwwq}
E_h^\beta  = 0. 
 \end{equation}
Therefore the ergotropy-to-battery-holder energy ratio for the linear quantum battery is, most pleasantly, a perfect one: $\mathcal{E}/E_h = 1$ [cf. Eq.~\eqref{eq:sfdhaaaarr}].

    \begin{figure*}[tb]
\begin{center}
 \includegraphics[width=1.0\linewidth]{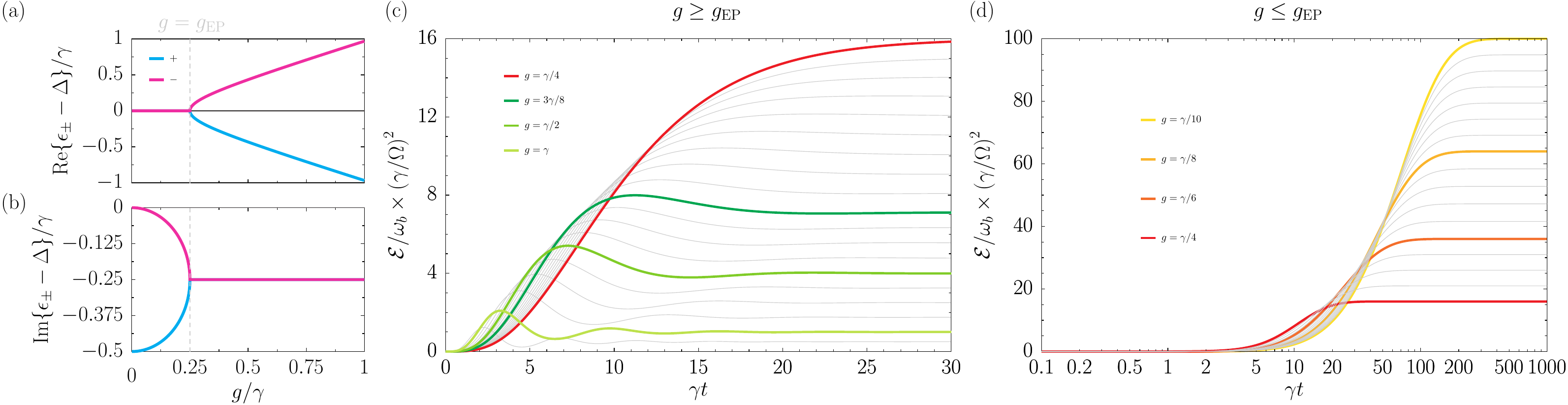}
 \caption{ \textbf{The linear quantum battery and its dynamics.} Panel (a): the real parts of the complex eigenvalues $\epsilon_\pm$, shifted by the drive-battery detuning $\Delta$, as a function of the charger-holder coupling strength $g$ (both in units of the charger damping rate $\gamma$) using Eq.~\eqref{eqapp:sdfazzsdffds}. Panel (b): as for panel (a), but for the imaginary parts. Dashed grey lines: the exceptional point $g_{\mathrm{EP}}$ [cf. Eq.~\eqref{eq:dfddfffsdfsfdsdfdfaa}]. Panel (c): the ergotropy $\mathcal{E}$, scaled by $(\gamma/\Omega)^2$ and in units of $\omega_b$, as a function of time $t$ (in units of the inverse of $\gamma$). Thick colored lines: the results for various values of the coupling $g$ above the exceptional point ($g \ge g_{\mathrm{EP}}$). Thin grey lines: intermediate couplings as guides for the eye. Panel (d): as for panel (c), but for couplings $g$ below the exceptional point ($g \le g_{\mathrm{EP}}$), and plotted as a semi-logarithmic plot. In panels (c) and (d), we consider the case of zero detuning ($\Delta = 0$), as governed by Eq.~\eqref{eq:q222}, Eq.~\eqref{eq:qdfdf222} and Eq.~\eqref{eq:qdfdfdfdf222}.  }
 \label{Figure2}
 \end{center}
\end{figure*}

In the steady state ($t \to \infty$), the ergotropy $\mathcal{E}$, and identically the energy $E_h$ of the battery holder, are simply given by [see Eq.~\eqref{eq:dpvrdfr} and Supplementary Note 2]
    \begin{equation}
    \label{eq:sfdsdfwwwq}
    \lim_{t \to \infty} \mathcal{E}  =   \lim_{t \to \infty} E_h  = \omega_b\frac{  \left( g \Omega \right)^2  }{ \left( \Delta^2 - g^2 \right)^2  + \left( \frac{ \gamma}{2} \Delta \right)^2 }. 
 \end{equation}
 Quite intuitively, maximising the steady state ergotropy $\mathcal{E}$ requires the maximisation of the driving amplitude $\Omega$ and the minimisation of the charger decay rate $\gamma$. There is a more complicated competition between the charger-holder coupling strength $g$ and the driving-battery detuning $\Delta$. Typical circumstances are plotted in Fig.~\ref{Figure1}~(b, c) for example cases with the couplings $g = \gamma$ and $g = \gamma/(2\sqrt{2}) \simeq 0.35 \gamma$ respectively. In these two panels, the darker the colour the larger the steady state ergotropy $\mathcal{E}$, showcasing how the ergotropic maximum depends nontrivially on the detuning $\Delta$. Denoting the specific value of the detuning $\Delta$ corresponding to the ergotropy maximum in the steady state as $\Delta'$, one finds the relation
      \begin{equation}
          \label{eq:dfdfdfdfaa}
\Delta'  =  \sqrt{g^2 - \left( \tfrac{\gamma}{2\sqrt{2}} \right)^2 } ~ \Theta \left(g - \tfrac{\gamma}{2\sqrt{2}} \right),
 \end{equation}
 where $\Theta (x)$ is the Heaviside step function. This formula suggests that zero detuning is optimal for smaller couplings $g \le \gamma /(2\sqrt{2})$ below a critical value, while otherwise a nonzero detuning is preferable -- as is hinted at in Fig.~\ref{Figure1}~(b, c) and as shown explicitly in Fig.~\ref{Figure1}~(d). Inserting Eq.~\eqref{eq:dfdfdfdfaa} into Eq.~\eqref{eq:sfdsdfwwwq} yields the most desirable steady state egrotopies $\mathcal{E}$, depending upon the relative weightings between $g$ and $\gamma$, as follows
       \begin{equation}
                 \label{eq:dfddfffdfdfaa}
\lim_{t \to \infty} \mathrm{max} \{ \mathcal{E} \} = \begin{cases}
  \omega_b \left( \frac{\Omega}{g} \right)^2,  & g \le \frac{\gamma}{2\sqrt{2}}, \\
 \omega_b \left( \frac{\Omega}{\gamma} \right)^2 \frac{\left( 4 g \right)^2}{\left( 2 g \right)^2-\left( \frac{\gamma}{2} \right)^2}, & g > \frac{\gamma}{2\sqrt{2}},
  \end{cases}
 \end{equation}
 as is plotted explicitly in Fig.~\ref{Figure1}~(e). Most notably, for smaller couplings $g \le \gamma /(2\sqrt{2})$ there is a simple quadratic relation for the largest steady state egrotropy, as was discussed in detail previously~\cite{Farina2019}. For larger couplings $g \gg \gamma /(2\sqrt{2})$, the largest steady state egrotopy becomes independent of the coupling $g$ as it tends towards $ \lim_{t \to \infty} \mathcal{E}  = \omega_b (2\Omega / \gamma)^2$, which features a natural competition between the drive amplitude $\Omega$ and the decay rate $\gamma$.

 The time-dependent behaviour of this linear quantum battery is mainly determined by the complex eigenvalues $\epsilon_{\pm}$, which arise from the dynamical matrix describing the first moments of the system (see Supplementary Note 2), like so
  \begin{equation}
\label{eqapp:sdfazzsdffds}
\epsilon_{\pm} = \Delta \pm G  - \mathrm{i} \frac{\gamma}{4},
\quad\quad\quad
\phi_{\pm} =
\begin{pmatrix}
1 \\
\frac{g}{\epsilon_{\pm} - \Delta}
\end{pmatrix},
 \end{equation}
where $\phi_{\pm}$ are the associated eigenvectors. Here we have introduced the renormalized coupling strength $G$ and (for use later on) a closely associated parameter known here as the renormalized decay rate $\Gamma$, which are defined together via 
    \begin{equation}
             \label{eq:dfdsdfsdfsdfsdfsdfffsdfsfdsdfdfaa}
G =\sqrt{ g^2 - \left( \tfrac{\gamma}{4} \right)^2  },
\quad\quad\quad
\Gamma =\sqrt{  \left( \tfrac{\gamma}{4} \right)^2 - g^2 }.
 \end{equation}
 The real and imaginary parts of the complex eigenvalues $\epsilon_{\pm}$ are plotted in Fig.~\ref{Figure2}~(a, b) using Eq.~\eqref{eqapp:sdfazzsdffds}. Most interestingly, when the coupling strength satisfies $g = g_{\mathrm{EP}}$ (dashed and grey vertical lines in both panels), where
    \begin{equation}
              \label{eq:dfddfffsdfsfdsdfdfaa}
g_{\mathrm{EP}} = \frac{\gamma}{4},
 \end{equation}
 the twin objects provided in Eq.~\eqref{eqapp:sdfazzsdffds} simultaneously coalesce, such that Eq.~\eqref{eq:dfddfffsdfsfdsdfdfaa} highlights the presence of an exceptional point in the dynamical system. Exceptional points are known to be highly consequential in systems which may be described as non-Hermitian in some sense, and typically they mark the borderland between regimes with very different physical responses~\cite{Berry2003, Heiss2003, Ozdemir2019, Miri2019,Downing2021}.

 The dynamical ergotropy $\mathcal{E}$ can be calculated using the framework built in Supplementary Note 2, and the analysis is most easily done for the case of zero drive-battery detuning ($\Delta = 0$). We start by noting that in the completely dissipationless battery charger limit (that is, when $\gamma \to 0$), one finds the oscillating ergotropy
\begin{equation}
\label{eq:ddfsddfgfgfsdffg}
  \lim_{\gamma \to 0}  \mathcal{E} = \omega_b \left( \frac{2 \Omega}{g} \right)^2 \sin^4 \left( \frac{g t}{2} \right),
\end{equation}
which clearly reaches its maximum value of $\omega_b (2\Omega/g)^2$ at the time $t = \pi/g$ (and at all later times with the periodicity of $2 \pi/g$). Within this lossless regime, we have also obtained the result equivalent to Eq.~\eqref{eq:ddfsddfgfgfsdffg} but with the counter-rotating terms in the coupling Hamiltonian of Eq.~\eqref{eq:sfsdfgsdghbnnfagttt} also included, which may be useful for future considerations of ultrastrongly coupled systems (see Supplementary Note 2). Otherwise, within the full driven-dissipative theory, there are two regimes of interest (split by an intermediate marginal case) due to the presence of the exceptional point of Eq.~\eqref{eq:dfddfffsdfsfdsdfdfaa}. These cases may be described analytically with the expressions
     \begin{widetext}
    \begin{align}
\mathcal{E} &= \omega_b \left( \frac{\Omega}{g} \right)^2 \Bigl\{ 1 - \bigr[ \cosh \left( \Gamma t \right) + \tfrac{\gamma}{4 \Gamma} \sinh \left( \Gamma t \right) \bigr] \mathrm{e}^{- \frac{\gamma}{4} t} \Bigl\}^2,   &&g < g_{\mathrm{EP}}, \label{eq:q222} \\
\mathcal{E} &= \omega_b  \left( \frac{4 \Omega}{\gamma} \right)^2 \Bigl\{ 1 - \bigr[ 1 + \tfrac{\gamma t}{4} \bigr] \mathrm{e}^{- \frac{\gamma}{4} t} \Bigl\}^2,  &&g = g_{\mathrm{EP}}, \label{eq:qdfdf222} \\
\mathcal{E} &= \omega_b \left( \frac{\Omega}{g} \right)^2 \Bigl\{ 1 - \bigr[ \cos \left( G t \right) + \tfrac{\gamma}{4 G} \sin \left( G t \right) \bigr]  \mathrm{e}^{- \frac{\gamma}{4} t} \Bigl\}^2,   &&g > g_{\mathrm{EP}},  \label{eq:qdfdfdfdf222}
 \end{align}
      \end{widetext}
      where the twin frequencies $G$ and $\Gamma$ are both defined in Eq.~\eqref{eq:dfdsdfsdfsdfsdfsdfffsdfsfdsdfdfaa}. Remarkably, the physics of the exceptional point ensures damped-hyperbolic, damped-algebraic and damped-trigonometric ergotropic behaviors are all possible, depending upon the relative strength of the charger-holder coupling $g$ as compared to the charger decay rate $\gamma$. Notably, we always maintain the coupling regime $g \ll \omega_b$ within our driven-dissipative theory in order to avoid the discarded counter-rotating terms in the original Hamiltonian becoming non-negligible, as was discussed after Eq.~\eqref{eq:lksdsad}. Nevertheless, in the extreme limiting cases of very weak ($g \ll \gamma$) and very strong ($g \gg \gamma$) couplings, as defined in comparison to the loss parameter $\gamma$ only, the ergotropy $\mathcal{E}$ follows the even simpler approximate expressions
   \begin{align}
\lim_{g \ll \gamma} \mathcal{E} &\simeq \omega_b \left( \frac{\Omega}{g} \right)^2 \Bigl\{ 1 - \mathrm{e}^{- \left( \frac{\gamma}{4} - \Gamma \right) t } \Bigl\}^2, \label{eq:dfgvxboo} \\
\lim_{g \gg \gamma} \mathcal{E} &= \omega_b \left( \frac{\Omega}{g} \right)^2 \Bigl\{ 1 - \cos (gt) \mathrm{e}^{- \frac{\gamma t}{4}} \Bigl\}^2.   \label{eq:dfgvxboo2}
 \end{align}
 With vanishing coupling $g$, the rough expression of Eq.~\eqref{eq:dfgvxboo} demonstrates the exponential rise in time of the erogtropy $\mathcal{E}$ towards its eventual steady state value [cf. Eq.~\eqref{eq:sfdsdfwwwq}]
 \begin{equation}
\label{eq:ddfdsfdfssddfgfgfsdffg}
  \lim_{t \to \infty}  \mathcal{E} = \omega_b \left( \frac{\Omega}{g} \right)^2,
\end{equation}
 as delicately controlled by the time constant $(\gamma/4-\Gamma)^{-1}$ and its double within Eq.~\eqref{eq:dfgvxboo}. The very strong coupling $g$ result of Eq.~\eqref{eq:dfgvxboo2} describes periodic oscillations, optimally bounded by $0$ and $\omega_b (2\Omega/g)^2$, which are gradually damped out towards the common steady state ergotropy of Eq.~\eqref{eq:ddfdsfdfssddfgfgfsdffg}, which is notably at a value of one quarter of its dynamical maximum.
 
   \begin{figure*}[tb]
\begin{center}
 \includegraphics[width=1.0\linewidth]{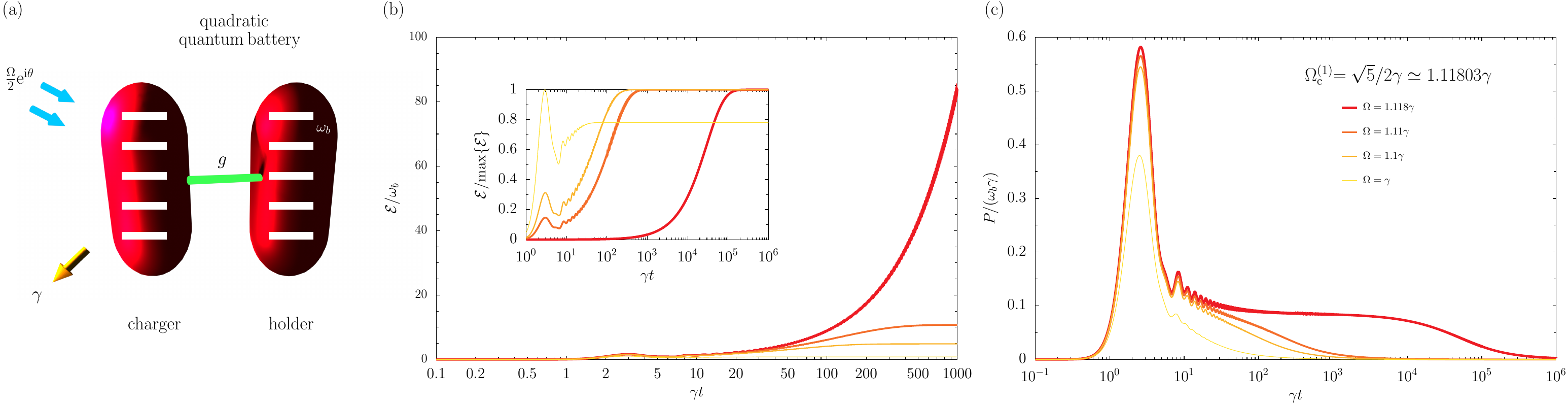}
 \caption{ \textbf{The quadratic quantum battery and its dynamics.} Panel (a): a sketch of the bipartite quantum battery (both parts are modelled as harmonic oscillators, with the level spacing $\omega_b$), composed of a battery charger coupled to a battery holder at the coupling strength $g$ (green bar). The charger is driven parametrically (two cyan arrows) with an amplitude $\Omega$, phase $\theta$ and frequency $2\omega_d$, while it suffers loss (yellow arrow) at the decay rate $\gamma$. Panel (b): a semi-logarithmic plot of the ergotropy $\mathcal{E}$  (in units of $\omega_b$) as a function of time $t$ (in units of the inverse of $\gamma$) [cf. Eq.~\eqref{eq:dpvrdfr} and Eq.~\eqref{eq:dpvrdfr2} with Eq.~\eqref{eqapp:sdfsdfsvsdfdsfesgrsgr}]. Inset: as for panel (b), but with the ergotropy $\mathcal{E}$ scaled by its maximum value, and with the timescale considered extended. Panel (c): a semi-logarithmic plot of the power $P = \mathcal{E}/t$  (in units of $\omega_b \gamma$) as a function of time $t$. In panels (b) and (c), we consider the case where $g=\gamma$ and $\Delta = \gamma/2$, so that from Eq.~\eqref{eq:sfdfgdfgsfdsf} the first critical driving amplitude $\Omega_{\mathrm{c}}^{(1)} = \sqrt{5}/2 \gamma \simeq 1.118 \gamma$, while the second critical driving amplitude $\Omega_{\mathrm{c}}^{(2)} = \sqrt{5/2} \gamma \simeq 1.581 \gamma$ [cf. Eq.~\eqref{eq:sfdfsdfdsfgdfgsfdsf}]. The plot legend in panel (c) also applies to panel (b), and marks the various driving amplitudes $\Omega$ considered.}
 \label{Figure3}
 \end{center}
\end{figure*}
 
The time-dependent ergotropy $\mathcal{E}$, for various couplings $g > g_{\mathrm{EP}}$ above the exceptional point, is shown in Fig.~\ref{Figure2}~(c) using the exact result of Eq.~\eqref{eq:qdfdfdfdf222}. This panel displays Rabi-style oscillations (which are stronger with larger $g$) before an eventual decay into a steady state with a rather moderate ergotropy $\mathcal{E}$. The red line in panel (c) represents the border case, as governed by Eq.~\eqref{eq:qdfdf222}, where the coupling $g = g_{\mathrm{EP}}$ exactly. For smaller couplings $g < g_{\mathrm{EP}}$ below the exceptional point, the dynamics is similarly plotted in Fig.~\ref{Figure2}~(d) using Eq.~\eqref{eq:q222}. Now Rabi-style oscillations are entirely absent, and the system instead monotonically increases in time towards a plateau at a much larger ergotropy $\mathcal{E}$ [cf. Eq.~\eqref{eq:ddfdsfdfssddfgfgfsdffg}], albeit at the cost of much larger charging times. We have checked that the general physics of the linear quantum battery, including the interesting exceptional point physics, is maintained even after the inclusion of loss from the battery holder (see Supplementary Note 2).
\\

%===========================================================================
%===========================================================================
%===========================================================================
%===========================================================================
%===========================================================================
%===========================================================================
%===========================================================================
%===========================================================================

\noindent \textit{Quadratic quantum battery}\\
The perspectives for a highly ergotropic quantum battery can arguably be improved by instead considering parametric driving~\cite{Minganti20166, Puri2017, Candia2023, Vidiella2023}. The quadratic nature of the resultant Hamiltonian ensures that the ergotropy-to-battery holder energy ratio can be readily calculated using the same Gaussian theoretical framework as for the linear quantum battery [cf. Eq.~\eqref{eq:dpvrdfr}, Eq.~\eqref{eq:dpvrdfr2} and Eq.~\eqref{eq:dpvrdfsfdsdfdfr}], while the two-photon driving should lead to larger population inversions.

\begin{figure*}[tb]
\begin{center}
 \includegraphics[width=1.0\linewidth]{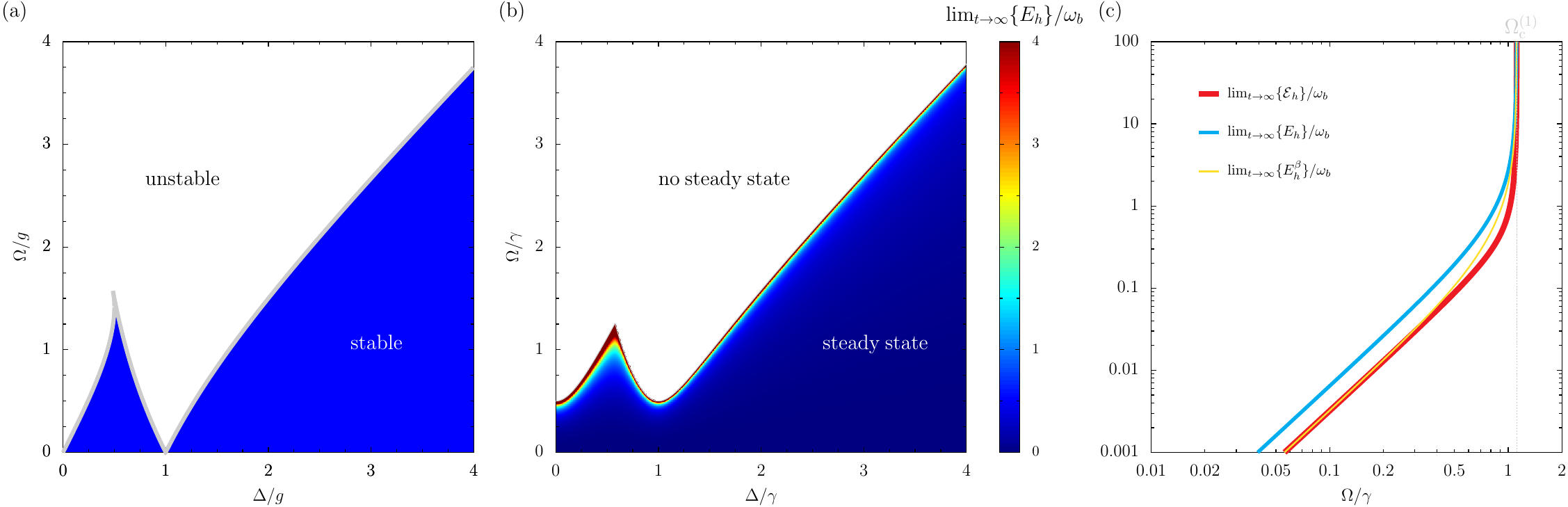}
 \caption{ \textbf{The quadratic quantum battery and its phase diagram.} Panel (a): the Hamiltonian phase diagram of the quantum battery, as a function of the drive-battery detuning $\Delta$ and the two-photon driving amplitude $\Omega$ (both in units of the charger-holder coupling strength $g$). The stable (blue) region is associated with wholly real Hamiltonian eigenfrequencies $\omega_{\pm}$, and the unstable (white) region corresponds to at least one complex eigenfrequency [cf. Eq.~\eqref{eq:sfdfgdsdssvvvfgsfdsf}]. Grey lines: boundaries coming from Eq.~\eqref{eq:dgbeeeededed}, Eq.~\eqref{eq:dgbeeeededed2} and Eq.~\eqref{eq:dgbeeeededed3}. Panel (b): the Liouvillian phase diagram of the quantum battery in the steady state ($t \to \infty$), as a function of $\Delta$ and $\Omega$ (both in units of the charger damping rate $\gamma$). The white region corresponds to the situation where no steady state can be formed, while the coloured region describes when a steady state is established. The borders are given by Eq.~\eqref{eq:sfdfgdfgsfdsf} and Eq.~\eqref{eq:sfdfsdfdsfgdfgsfdsf}. Color bar: the battery holder energy $E_h$ in the steady state (in units of the battery holder energy level spacing $\omega_b$) from Eq.~\eqref{eq:dgfdgdgz}. In this panel, we consider the case of $g = \gamma$, and we cap the energy at the maximum value of $E_h = 4 \omega_b$ in the color bar. Panel (c): a logarithmic plot of the salient steady state energies as a function of $\Omega$ (in units of $\gamma$). The ergotropy $\mathcal{E}$ (thick red line) [cf. Eq.~\eqref{eq:sfdhaaaarr}], the battery holder energy $E_h$ (medium cyan line) [cf. Eq.~\eqref{eq:dgfdgdgz}], and the battery holder energy in the passive state $E_h^\beta$ (thin yellow line) [cf. Eq.~\eqref{eq:dpvrdfr2} with Eq.~\eqref{eqapp:sdfhgjhsdfsdfssvfesgrsgr}] are all shown. Dashed grey line: the critical driving amplitude $\Omega_{\mathrm{c}}^{(1)}$. In this panel, we consider $\Delta = \gamma/2$ and $g = \gamma$, so that $\Omega_{\mathrm{c}}^{(1)} = \sqrt{5} \gamma/2 \simeq 1.118 \gamma$ and $\Omega_{\mathrm{c}}^{(2)} = \sqrt{5/2} \gamma \simeq 1.581 \gamma$ from Eq.~\eqref{eq:sfdfgdfgsfdsf} and Eq.~\eqref{eq:sfdfsdfdsfgdfgsfdsf} respectively.    }
 \label{Figure4}
 \end{center}
\end{figure*}

Let us replace the original driving Hamiltonian $\hat{H}_{d}$ of Eq.~\eqref{eq:lksdsad} with a two-photon drive given by
   \begin{equation}
\hat{H}_{d} = \frac{\Omega}{2} \left( \mathrm{e}^{\mathrm{i} \theta} \mathrm{e}^{-2\mathrm{i} \omega_d t} c^\dagger c^\dagger + \mathrm{e}^{-\mathrm{i} \theta} \mathrm{e}^{2\mathrm{i} \omega_d t} c c \right), 
 \end{equation}
 which nicely fits into the full Hamiltonian operator $\hat{H}$ of Eq.~\eqref{eq:Haxcsdfdsfdsfsdfvcxvmy}. This quadratic quantum battery model is sketched in Fig.~\ref{Figure3}~(a), where the parametric drive is of amplitude $\Omega \ge 0$, phase $\theta$ and frequency $2\omega_d$. The analysis leading to the quantum master equation of Eq.~\eqref{eq:mnnbfgvd} also holds for the quadratic driving case, after the replacement of the effective Hamiltonian operator $ \hat{\mathcal{H}}$ of Eq.~\eqref{eq:mnnbfgvd2} with
\begin{align}
\label{eq:Haxcsdfsdfvcxvmy}
 \hat{\mathcal{H}} =~& \Delta \left( c^{\dagger} c + h^{\dagger} h \right) + g \left( c^{\dagger} h + h^{\dagger} c \right) \nonumber \\
 &+ \frac{\Omega}{2} \mathrm{e}^{\mathrm{i} \theta} c^{\dagger} c^{\dagger} +  \frac{\Omega}{2} \mathrm{e}^{-\mathrm{i} \theta} c c.
\end{align}
The two-photon driving appearing within Eq.~\eqref{eq:Haxcsdfsdfvcxvmy} implies the presence of quantum squeezing~\cite{Loudon1987}, in stark contrast to the one-photon driving case described previously in Eqs.~\eqref{eq:sfsfagttt}--\eqref{eq:lksdsad}. The quadratic Hamiltonian operator $\hat{\mathcal{H}}$ provided in Eq.~\eqref{eq:Haxcsdfsdfvcxvmy} may be diagonalized exactly by bosonic Bogoliubov transformation~\cite{Tsallis1978, Colpa1978} into the two-mode form
\begin{equation}
\label{eq:sfsfdsf}
 \hat{\mathcal{H}} = \sum_{\tau = \pm} \omega_{\tau} \beta_{\tau}^\dagger \beta_{\tau},
\end{equation}
where the mode index $\tau = \pm$ characterizes the two Bogoliubov eigenfrequencies $\omega_{\pm}$, as defined by the expression
\begin{equation}
\label{eq:sfdfgdsdssvvvfgsfdsf}
  \omega_{\pm} = \sqrt{ \Delta^2+g^2 - \tfrac{\Omega^2}{2} \pm \sqrt{  \Omega^2 \left( \tfrac{\Omega^2}{4} - g^2 \right)  + 4 g^2 \Delta^2 }}. 
\end{equation}
The twin Bogoliubov operators $\beta_{\pm}$ appearing in Eq.~\eqref{eq:sfsfdsf} satisfy the standard bosonic commutation rule $[ \beta_{\pm},  \beta_{\pm}^\dagger] = 1$, and they are composed of all four operators of the problem ($c$, $h$, $c^\dagger$, and $h^\dagger$) as follows
\begin{equation}
\label{eq:sfdffgdgdfgsfdsf}
 \beta_{\pm} = \frac{1}{\sqrt{2}}
 \begin{pmatrix}
\cosh \left( \mu_{\pm} \right) \\
\cosh \left( \nu_{\pm} \right)  \\
\mathrm{e}^{\mathrm{i} \theta} \sinh \left( \mu_{\pm} \right)  \\
\mathrm{e}^{\mathrm{i} \theta} \sinh \left( \nu_{\pm} \right)
\end{pmatrix}^{\mathrm{T}} 
\cdot
 \begin{pmatrix}
c \\
h \\
c^\dagger \\
h^\dagger
\end{pmatrix}, 
\end{equation}
where the Bogoliubov coefficients in Eq.~\eqref{eq:sfdffgdgdfgsfdsf} may be found from the hyperbolic tangent relations
\begin{align}
\label{eq:sfdffdsfsgdgdfgsfdsf}
\tanh \left( \mu_{\pm} \right) &= \frac{ \Omega \left( \Delta + \omega_\pm \right) }{ \left( \Delta + \omega_\pm \right)^2 - g^2 }, \\
\tanh \left( \nu_{\pm} \right) &=  \frac{ \Omega \left( \Delta - \omega_\pm \right) }{ \left( \Delta + \omega_\pm \right)^2 - g^2 }.
\end{align}
We also note that inverting the operators of Eq.~\eqref{eq:sfdffgdgdfgsfdsf} allows for the nontrivial vacuum state populations of the battery charger and battery holder to be calculated (see Supplementary Note 3), while the squeezing promised by the counter-rotating driving terms is considered later on. Most interestingly, the two Bogoliubov eigenfrequencies $\omega_{\pm}$ of Eq.~\eqref{eq:sfdfgdsdssvvvfgsfdsf} are not wholly real for all values of the three Hamiltonian parameters $\Delta$, $\Omega$ and $g$. This suggests a spectral collapse~\cite{Kirton2018, Salado2021} within the purely Hamiltonian theory of Eq.~\eqref{eq:Haxcsdfsdfvcxvmy} in certain parameter regimes. The phase diagram implied by the stability of the solely Hamiltonian operator $\hat{\mathcal{H}}$ approach is plotted in Fig.~\ref{Figure4}~(a), where the boundaries (grey lines) are defined by the three equations [cf. when Eq.~\eqref{eq:sfdfgdsdssvvvfgsfdsf} become complex]
 \begin{align}
\label{eq:dgbeeeededed}
\Omega &= \Big| \Delta - \tfrac{g^2}{\Delta} \Big| \Theta \left( \tfrac{\Delta}{g} - \Xi \right), \\
   \Omega &= g \left( \sqrt{1 + 2 \tfrac{\Delta}{g}  } - \sqrt{1 - 2 \tfrac{\Delta}{g}} \right) \Theta \left( \tfrac{1}{2} - \tfrac{\Delta}{g} \right),  \label{eq:dgbeeeededed2} \\
      \Omega &= g \left( \sqrt{1 + 2 \tfrac{\Delta}{g}  } + \sqrt{1 - 2 \tfrac{\Delta}{g}} \right) \Theta \left( \tfrac{1}{2} - \tfrac{\Delta}{g} \right)  \Theta \left( \tfrac{\Delta}{g} - \Xi \right),  \label{eq:dgbeeeededed3} 
\end{align} 
where the dimensionless number $\Xi = \sqrt{ \sqrt{5} - 2 } \simeq 0.486$. The diagram of Fig.~\ref{Figure4}~(a) hints at a classification where the stable regions (blue, real $\omega_{\pm}$) are associated with convergent-in-time dynamics, while the unstable regions (white, complex $\omega_{\pm}$) should not be able to support a steady state due to their divergent-in-time dynamics. The introduction of dissipation into the system via the quantum master equation of Eq.~\eqref{eq:mnnbfgvd} upgrades the simple Hamiltonian dynamics of Eq.~\eqref{eq:sfsfdsf} and necessarily updates the phase diagram of Fig.~\ref{Figure4}~(a) in a more physically meaningful manner (for example, the imaginary parts of complex eigenfrequencies can then be interpreted as being related to inverse lifetimes), as we now explore with proper reference to the steady state of the full driven-dissipative system.

Within an open quantum systems approach, the steady state energy of the battery holder $E_h$ [cf. Eq.~\eqref{eq:dpvrdfr}] may be found from the first and second moments of the system (see Supplementary Note 3 for the full theory). This leads to the analytic expression 
  \begin{widetext}
  \begin{equation}
  \label{eq:dgfdgdgz}
\lim_{t \to \infty } E_h  = \omega_b \frac{ \Omega^2  }{ \left( 2\Delta \right)^2 + \left( \frac{\gamma}{2} \right)^2 - \Omega^2 } \frac{  \Delta^4 + \frac{g^4}{2} - \Delta^2 \left[ \frac{g^2}{2} + \Omega^2 - \left( \frac{\gamma}{2} \right)^2 \right]  }{ \Delta^4 + g^4 - \Delta^2 \left[ 2 g^2 + \Omega^2 - \left( \frac{\gamma}{2} \right)^2 \right] }, 
 \end{equation}
   \end{widetext}
which may be compared to Eq.~\eqref{eq:sfdsdfwwwq}, the analogous result with one-photon driving. Notably, the denominators of both the first and second terms in the product of fractions comprising Eq.~\eqref{eq:dgfdgdgz} may each become zero at some point in parameter space, suggesting two critical driving amplitudes of the system -- which we call $\Omega_{\mathrm{c}}^{(1)}$ and $ \Omega_{\mathrm{c}}^{(2)}$. These critical points are given by [cf. Eq.~\eqref{eq:dgbeeeededed}--\eqref{eq:dgbeeeededed3} from the Hamiltonian theory]
 \begin{align}
\label{eq:sfdfgdfgsfdsf}
  \Omega_{\mathrm{c}}^{(1)} &= \sqrt{ \left( \tfrac{\gamma}{2} \right)^2 + \left( 2 \Delta\right)^2  }, \\
    \Omega_{\mathrm{c}}^{(2)} &= \sqrt{  \left( \tfrac{\gamma}{2} \right)^2 + \Delta^2 \left( 1 - \tfrac{g^2}{\Delta^2} \right)^2 },  \label{eq:sfdfsdfdsfgdfgsfdsf}
\end{align} 
and help to define the Liouvillian phase diagram of the quadratic quantum battery in the steady state, via the duo of equations $\Omega = \Omega_{\mathrm{c}}^{(1)} \Theta ( 1/ \sqrt{3} - \Delta/g )$ and $\Omega = \Omega_{\mathrm{c}}^{(2)} \Theta ( \Delta/g - 1/ \sqrt{3} )$. The steady state battery holder energy $E_h$ is plotted in Fig.~\ref{Figure4}~(b), where the situation without a steady state is represented by the white region. Clearly, approaching the critical line formed using Eq.~\eqref{eq:sfdfgdfgsfdsf} and Eq.~\eqref{eq:sfdfsdfdsfgdfgsfdsf} leads to a transition in the steady state response of the quantum battery -- above this dynamical instability the energetics are unbounded since the mean battery populations are themselves unbounded. This is because the energetic drive into the battery more than compensates the loss into the external environment, leading to a dramatic rise of bosonic excitations up the infinite energy ladders comprising the quadratic quantum battery. The energy formula of Eq.~\eqref{eq:dgfdgdgz} becomes much simpler in the two limiting cases of small detuning ($\Delta \to 0$) and small battery charger-holder coupling ($g \to 0$), where
  \begin{align}
  \label{eq:bosds}
\lim_{t \to \infty } E_h  &= \frac{\omega_b}{2} \frac{ \Omega^2  }{  \left( \frac{\gamma}{2} \right)^2 - \Omega^2 }, \quad &&\Delta \to 0, \\
\lim_{t \to \infty } E_h  &= \omega_b \frac{ \Omega^2  }{  \left( 2\Delta \right)^2 + \left( \frac{\gamma}{2} \right)^2 - \Omega^2 }, \quad &&g \to 0.   \label{eq:bosds2}
 \end{align}
 The left-hand vertical axis of Fig.~\ref{Figure4}~(b) is explained by Eq.~\eqref{eq:bosds}, complete with its finishing point at $\Omega = \gamma/2$, above which no steady state is formed. Meanwhile, Eq.~\eqref{eq:bosds2} describes the weakly coupled battery result (not shown in Fig.~\ref{Figure4}~(b), which takes $g = \gamma$ as an example case) which sees a reduction in the number of critical points from two to one, located at $\Omega_{\mathrm{c}}^{(1)}$ only [cf. Eq.~\eqref{eq:sfdfgdfgsfdsf}]. The phase diagram of Fig.~\ref{Figure4}~(b) illustrates the critical nature of the quadratic quantum battery, where significant energies can be obtained near to dynamical instabilities governed by critical points, which may be starkly contrasted to the linear quantum battery which is instead dominated by exceptional point physics.

 The quadratic nature of the Hamiltonian operator $\hat{\mathcal{H}}$ given by Eq.~\eqref{eq:Haxcsdfsdfvcxvmy} ensures that the result of Eq.~\eqref{eq:dpvrdfr2}, quantifying the energy of the battery holder in the passive state $E_h^\beta$, also holds. In particular, the boundary conditions of the system imply that the first moments of the system are all zero (see Supplementary Note 3 for details). Therefore that the key quantity $\mathcal{D}$, as defined in Eq.~\eqref{eq:dpvrdfsfdsdfdfr}, reduces to the solely second moments form [cf. Eq.~\eqref{eq:sfdsdfwrfgsfdwwq} for the linear quantum battery]
   \begin{equation}
\label{eqapp:sdfsdfsvsdfdsfesgrsgr}
 \mathcal{D} =  \left( 1 + 2 \langle h^\dagger h \rangle \right)^2 - 4  \langle h^\dagger h^\dagger  \rangle  \langle h h  \rangle.
 \end{equation}
Notably, since in general $\mathcal{D} \ne 1$ the quadratic quantum battery is associated with a nonzero passive state energy ($E_h^\beta \ne 0$), which acts to reduce the ergotropy $\mathcal{E}$ following the definition of Eq.~\eqref{eq:sfdhaaaarr}. In the crucial steady state regime ($t \to \infty$), the explicit form of Eq.~\eqref{eqapp:sdfsdfsvsdfdsfesgrsgr} is derivable exactly as (see Supplementary Note 3)
 \begin{widetext}
   \begin{equation}
\label{eqapp:sdfhgjhsdfsdfssvfesgrsgr}
\fixTABwidth{T}
\scalemath{0.82}{ \lim_{t \to \infty} \mathcal{D} = \frac{ \left[ \left( 2 \Delta \right)^2 + \left( \frac{\gamma}{2} \right)^2 \right]^2 \left[ \left( \Delta^2 - g^2 \right)^2 + \Delta^2 \left( \frac{\gamma}{2} \right)^2 \right] - \Omega^2 \left[ \left( 2 \Delta \right)^2 + \left( \frac{\gamma}{2} \right)^2 \right] \left[ 2 \Delta^4 + g^4 - \Delta^2 \left( 2 g^2 + \left( \frac{\gamma}{2} \right)^2 \right) \right]  - \Delta^2 \Omega^4 \left[ 7 \Delta^2 - 4 g^2 + \left( \frac{\gamma}{2} \right)^2 \right] - \Delta^2 \Omega^6 }{  \left[ \left( 2\Delta \right)^2 + \left( \frac{\gamma}{2} \right)^2 - \Omega^2 \right]^2 \left[ \Delta^4 - \Delta^2 \left[ 2 g^2 + \Omega^2 - \left( \frac{\gamma}{2} \right)^2 \right] + g^4 \right]}}.
 \end{equation}
 \end{widetext}
 This analytic result ensures that the steady state ergotropy $\mathcal{E}$ has been analytically obtained with the aid of the analytic expression of the holder energy $E_h$ [cf. Eq.~\eqref{eq:dgfdgdgz}] alongside the exact passive state holder energy $E_h^\beta$ [cf. Eq.~\eqref{eq:dpvrdfr2} with Eq.~\eqref{eqapp:sdfhgjhsdfsdfssvfesgrsgr}]. The full expression for the steady state ergotropy $\mathcal{E}$ is particularly compact in the limiting case of small detuning ($\Delta \to 0$), where we find it reduces to
   \begin{equation}
  \label{eq:bossdfsdfdfsds}
\lim_{t \to \infty } \mathcal{E}  =  \frac{\omega_b}{2} \left( \frac{\left( \frac{\gamma}{2} \right)^2}{ \left( \frac{\gamma}{2} \right)^2 - \Omega^2} - \sqrt{ \frac{\left( \frac{\gamma}{2} \right)^2}{ \left( \frac{\gamma}{2} \right)^2 - \Omega^2} } \right), 
 \end{equation}
which exhibits a single critical point at $\Omega_{\mathrm{c}}^{(1)} = \gamma/2$ [cf. Eq.~\eqref{eq:sfdfgdfgsfdsf}]. In the limit of small charger-holder coupling $g$, one finds that the full expression reduces to the intuitive result of zero steady state ergotropy ($\lim_{t \to \infty } \mathcal{E}  = 0$ for the case of $g \to 0$).

We consider the steady state energetics of the quadratic quantum battery in Fig.~\ref{Figure4}~(c), as a function of the driving amplitude $\Omega$, for the example case where the detuning $\Delta = \gamma/2$ and the charger-holder coupling $g = \gamma$. The ergotropy $\mathcal{E}$ (thick red line) [cf. Eq.~\eqref{eq:sfdhaaaarr}], the battery holder energy $E_h$ (medium cyan line) [cf. Eq.~\eqref{eq:dgfdgdgz}], and the battery holder energy in the passive state $E_h^\beta$ (thin yellow line) [cf. Eq.~\eqref{eq:dpvrdfr2} with Eq.~\eqref{eqapp:sdfhgjhsdfsdfssvfesgrsgr}] are all shown explicitly in their steady state forms. As the driving amplitude $\Omega$ approaches the critical point at $\Omega_{\mathrm{c}}^{(1)} = \sqrt{5/2} \gamma \simeq 1.118 \gamma$ (dashed grey line) [cf. Eq.~\eqref{eq:sfdfgdfgsfdsf}] all three energetic quantities increase without bound, leading to an abundance of useful energy being stored in the battery. This boundlessness occurs since a steady state is no longer supportable within this driven-dissipative theory when the drive into the battery overcompensates the loss into the outside environment. In practice, such a seemingly runaway solution may be prevented by either truncating the infinite energy ladders associated with the harmonic oscillators forming the quantum battery model, leading instead to a saturation at some large energy (as is discussed later on), or by introducing anharmonicities. Importantly, there is no analogous critical points in the linear quantum battery, and hence there is no such optimal driving associated with a huge ergotropic response.

      \begin{figure*}[tb]
\begin{center}
 \includegraphics[width=1.0\linewidth]{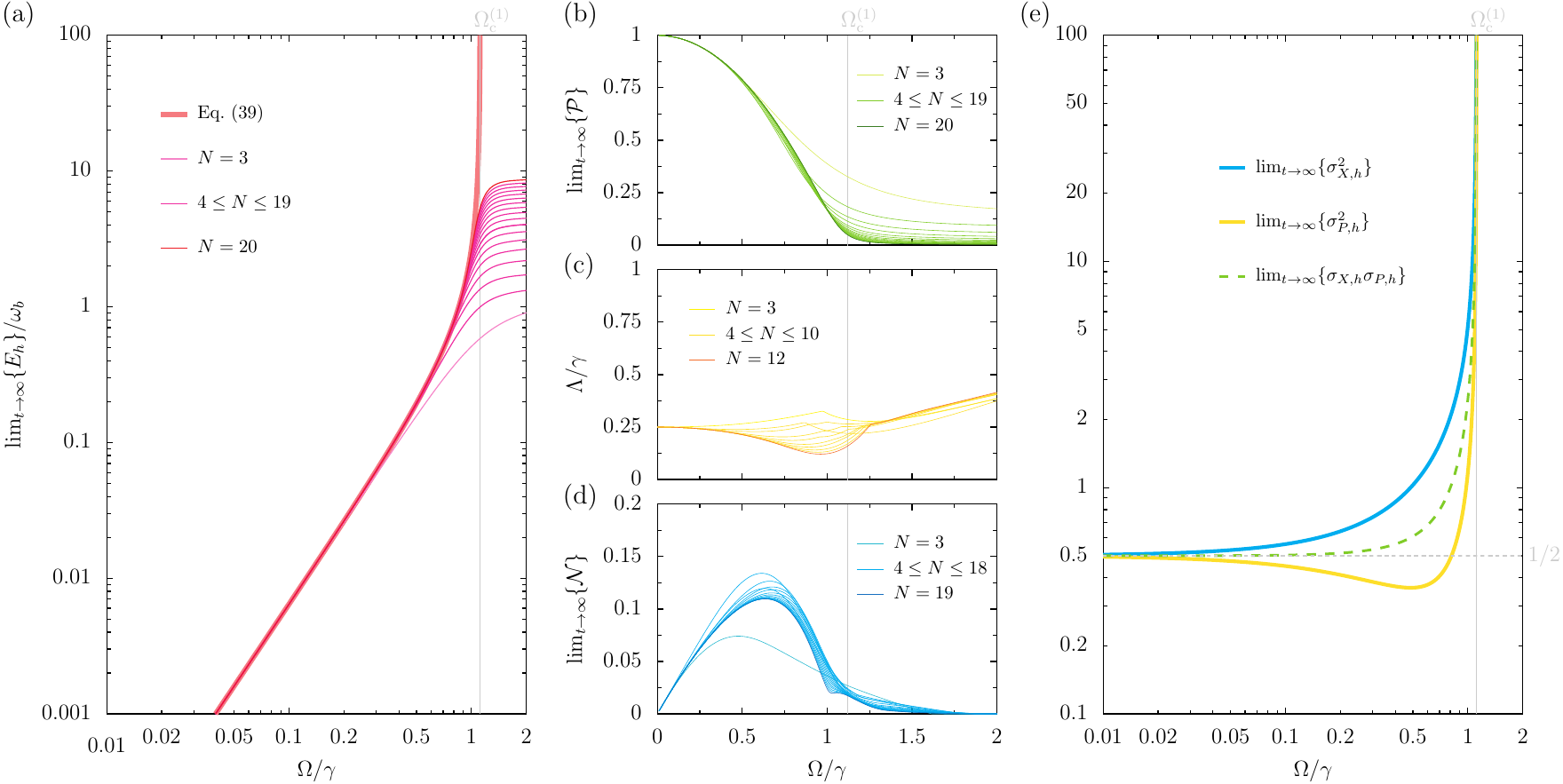}
 \caption{ \textbf{The quadratic quantum battery and its steady state.} Panel (a): a logarithmic plot of the battery holder energy $E_h$ in the steady state (in units of the holder energy level spacing $\omega_b$) as a function of the two-photon driving amplitude $\Omega$ (in units of the charger decay rate $\gamma$). Thick line: the analytic result of Eq.~\eqref{eq:dgfdgdgz}. Thin lines: results obtained by truncating the harmonic oscillators modelling the battery to $N$-level systems. Panel (b): the purity $\mathcal{P}$, in the steady state and as a function of $\Omega$, for a battery comprised of $N$-level systems. Panel (c): the Liouvillian gap $\Lambda$ as a function of $\Omega$. Panel (d): the negativity $\mathcal{N}$, in the steady state and as a function of $\Omega$. Panel (e): a logarithmic plot of the battery holder quadrature variances $\sigma_{X, h}^2$ (thick cyan line) and $\sigma_{P, h}^2$ (thick yellow line) in the steady state and as a function of $\Omega$. The product of the standard deviations $\sigma_{X, h} \sigma_{P, h}$ (medium dashed green line) is also shown, as is a guide for the eye at the Robertson-Schr\"{o}dinger minimum uncertainty of $1/2$ (horizontal, dashed grey line). In the figure, the charger-holder coupling strength $g=\gamma$ and the drive-battery detuning $\Delta = \gamma/2$, so that from Eq.~\eqref{eq:sfdfgdfgsfdsf} the first critical driving amplitude $\Omega_{\mathrm{c}}^{(1)} = \sqrt{5} \gamma/2 \simeq 1.118 \gamma$ (vertical, solid grey lines in all panels).    }
 \label{Figure5}
 \end{center}
\end{figure*}

The time-dependent properties of the ergotropy $\mathcal{E}$ are considered in Fig.~\ref{Figure3}~(b). There we again take the example quadratic quantum battery with the charger-holder coupling $g=\gamma$ and the detuning $\Delta = \gamma/2$, so that from Eq.~\eqref{eq:sfdfgdfgsfdsf} the first critical driving amplitude $\Omega_{\mathrm{c}}^{(1)} \simeq 1.118 \gamma$, while the second critical driving amplitude from Eq.~\eqref{eq:sfdfsdfdsfgdfgsfdsf} is $\Omega_{\mathrm{c}}^{(2)} \simeq 1.581 \gamma$. Increasingly strong driving amplitudes $\Omega$ are considered up to $\Omega_{\mathrm{c}}^{(1)}$ [see the plot legend in Fig.~\ref{Figure3}~(c) for the cases considered], leading to dramatic ergotropic improvements at larger timescales $\gamma t \gg 1$, as foreshadowed by the steady state analysis provided in Fig.~\ref{Figure4}~(c). The inset of Fig.~\ref{Figure3}~(b) rescales the ergotropy $\mathcal{E}$ by its maximum value over all times, showing the typical timescales required for the battery to become fully charged (in an ergotropic sense). Notably, full charging is achieved asymptotically ($t \to \infty$) except for when the driving amplitude is furthest away (thin yellow line) from $\Omega_{\mathrm{c}}^{(1)}$, where a dynamical ergotropic value is higher than its steady state value [a circumstance which commonly occurs for the linear quantum battery, see Fig.~\ref{Figure2}~(c)]. Finally, the equivalent charging powers $P = \mathcal{E}/t$ for the ergotropies considered in Fig.~\ref{Figure3}~(b) are shown in Fig.~\ref{Figure3}~(c). This panel~(c) shows that the power $P$ increases with time from zero up to some maximum power, which is seen to occur around the optimal time $t \simeq 2.6/\gamma$ for the cases considered. Hence the maximum power does not correspond to when the battery is fully charged [in the sense of the inset to Fig.~\ref{Figure3}~(b)]. A logarithmic timescale is used in Fig.~\ref{Figure3}~(c) to confirm the eventual loss of power when the steady state ergotropy is finally reached at large timescales. Notably, we have made sure that the general physics of the quadratic quantum battery, as described in Fig.~\ref{Figure3} and Fig.~\ref{Figure4} in particular, is essentially unchanged after including additional dissipation coming from the battery holder itself (see Supplementary Note 3).

The steady state properties of the quadratic quantum battery are further considered in Fig.~\ref{Figure5}. In the figure, the charger-holder coupling strength remains at $g=\gamma$ and the drive-battery detuning remains at $\Delta = \gamma/2$, so that the critical point $\Omega_{\mathrm{c}}^{(1)} \simeq 1.118 \gamma$ (vertical, solid grey lines in all panels of Fig.~\ref{Figure5}). In particular, we are interested in the results obtained by truncating the two harmonic oscillators modelling the battery charger and the battery holder to finite $N$-level systems, in order to better understand where the energetic unboundedness suggested in Fig.~\ref{Figure4}~(c) arises from.

In Fig.~\ref{Figure5}~(a), we consider the battery holder energy $E_h$ in the steady state (in units of the battery energy level spacing $\omega_b$) as a function of the two-photon driving amplitude $\Omega$ (in units of the charger decay rate $\gamma$). The analytic (and $N \to \infty$ asymptotic) result of Eq.~\eqref{eq:dgfdgdgz} is represented by the thick salmon-pink line, showcasing the aforementioned unboundedness as $\Omega \to \Omega_{\mathrm{c}}^{(1)}$. Otherwise, the truncated oscillator results are denoted by thin pink lines (the extreme cases for $N = 3$ and $N = 20$ levels are distinguished with light pink and red lines respectively). The data in Fig.~\ref{Figure5}~(a) confirms the increasing impact of the first critical driving amplitude $\Omega_{\mathrm{c}}^{(1)}$ with an increasing number of levels $N$, and highlights the eventual plateauing of the battery holder energy $E_h$ due to the necessary saturation of the finite level systems making up the truncated battery. These results support the main findings presented in Fig.~\ref{Figure4}~(c) for infinite-level quantum harmonic oscillators, and imply that the energetic unboundedness is a result of always being able to occupy higher and higher levels of an untruncated harmonic oscillator, as opposed to occurring due to an unphysical runaway or some unreasonable approximation.

The behavior of the purity $\mathcal{P}$ of the quadratic quantum battery, a measure of the degree of mixedness of the quantum state $\rho$ via the formula $\mathcal{P} = \Tr (\rho^2)$, is likewise shown in Fig.~\ref{Figure5}~(b) for the steady state. The purity is bounded by $\mathcal{P} =1$ for a pure state and $\mathcal{P} = 1/d$ (where $d$ is the dimension of the relevant Hilbert space) for a maximally mixed state. With vanishing driving $\Omega$, the system remains in its vacuum state and is hence completely pure. However for non-vanishing driving $\Omega$ the key quantity $\mathcal{D}$ [cf. Eq.~\eqref{eq:dpvrdfsfdsdfdfr}] is non-unity, and the degree of mixedness rapidly increases as the driving amplitude $\Omega$ is increased towards the critical point $\Omega_{\mathrm{c}}^{(1)}$, eventually leading to a maximally mixed quantum state and significant loss of coherences. This impure behaviour is in stark contrast to the completely pure linear quantum battery, which remains in a product state such that it exhibits the property $\mathcal{D} = 1$.

The underlying nature of the critical driving amplitude $\Omega_{\mathrm{c}}^{(1)}$ can be revealed by considering the response of the Liouvillian gap $\Lambda$. This important quantity is defined as the gap between zero and the real part of the largest (nonzero) eigenvalue in the Liouvillian superoperator spectrum~\cite{Kessler2012, Cai2013, Minganti2018}, and is displayed in Fig.~\ref{Figure5}~(c). The closing of the Liouvillian gap (which is possible in the truly thermodynamic limit, where $N \to \infty$) at some critical value of a system parameter is typically associated with a dissipative phase transition~\cite{Fitzpatrick2017, Rodriguez2017, Fink2018}. In the case of the quadratic quantum battery, the crucial parameter again seems to be the critical driving amplitude $\Omega_{\mathrm{c}}^{(1)}$ (more evidence supporting this conjecture is provided in Supplementary Note 3, which suggests an algebraic scaling of the Liouvillian gap size with the system size $N$). No such closing of the Liouvillian gap is possible for the one-photon driving case considered previously for the linear quantum battery, since in that case there is always a well-defined steady state [cf. Eq.~\eqref{eq:sfdsdfwwwq} with Eq.~\eqref{eq:dgfdgdgz}] and as such no dynamical instability is present. 

The fact that the key quantity $\mathcal{D} \ne 1$ [cf. Eq.~\eqref{eq:dpvrdfsfdsdfdfr}] for the quadratic quantum battery raises the possibility that quantum entanglement is playing a role. To investigate this, we consider the negativity $\mathcal{N}$, a common entanglement measure defined using the absolute sum of the negative eigenvalues of the partial transpose of the density matrix $\rho$~\cite{Zyczkowski1998, Vidal2002}. Zero negativity implies an unentangled state, while increasingly nonzero negativity suggests an increasingly entangled state. We plot the negativity $\mathcal{N}$, again in the steady state and as a function of the driving amplitude $\Omega$, in Fig.~\ref{Figure5}~(d). The plot shows that the state is completely unentangled with vanishing driving $\Omega \to 0$, since it is simply the trivial vacuum state. However, with increasingly strong driving $\Omega$ the entanglement of the state rises to a certain maximum, before falling once again towards zero (in the thermodynamic $N \to \infty$ limit) near to the critical point $\Omega_{\mathrm{c}}^{(1)}$. This trend occurs since, for any truncated oscillator case, the state at some large enough driving $\Omega$ is simply that with the highest level filled. This type of entanglement behaviour is entirely missing in the linear quantum battery, which always remains in an unentangled product state. 

Finally, the phenomena of quantum squeezing within the quadratic quantum battery may be analyzed through the two battery holder quadrature variances $\sigma_{X, h}^2$ and $\sigma_{P, h}^2$~\cite{Loudon1987}. These dimensionless quantities, defined via the twin relations of $\sigma_{X, h}^2 = \langle \hat{X}_h^2 \rangle - \langle \hat{X}_h \rangle^2$ (thick cyan line) and $\sigma_{P, h}^2 = \langle \hat{P}_h^2 \rangle - \langle \hat{P}_h \rangle^2$ (thick yellow line), are considered (in the steady state and as a function of the driving amplitude $\Omega$) in Fig.~\ref{Figure5}~(e). These variance definitions rely on the generalized battery holder quadratures $\hat{X}_h = ( \mathrm{e}^{\mathrm{i} \theta /2} h^\dagger +  \mathrm{e}^{-\mathrm{i} \theta/2} h) /\sqrt{2}$ and $\hat{P}_h = \mathrm{i} ( \mathrm{e}^{\mathrm{i} \theta/2 } h^\dagger -  \mathrm{e}^{-\mathrm{i} \theta /2} h) / \sqrt{2}$, which obey the familiar commutation relation $[ \hat{X}_h, \hat{P}_h] = \mathrm{i}$. The product of the standard deviations $\sigma_{X, h} \sigma_{P, h}$ (medium dashed green line) is also shown in Fig.~\ref{Figure5}~(e), as is a guide for the eye at the Robertson-Schr\"{o}dinger minimum uncertainty of $1/2$ (horizontal, dashed grey line). Most notably, the quasi-momentum variance $\sigma_{P, h}^2$ displays steady state squeezing for a range of driving amplitudes $\Omega$, up to a certain value $\Omega \simeq 0.813 \gamma$ (which is notably below the critical point residing at $\Omega_{\mathrm{c}}^{(1)} \simeq 1.118 \gamma$). Such quadrature squeezing of the quadratic quantum battery directly originates from the parametric field driving and thus is entirely absent for the coherent field driving case of the linear quantum battery. Squeezing may be interesting for modern applications in quantum sensing and for quantum information processing, while here it is interesting to note that the quadrature variances display asymptotes at $\Omega_{\mathrm{c}}^{(1)}$, the onset of the dynamical instability discussed earlier. Within wider quantum battery research, squeezing has been recently studied in the context of a coherent squeezing charging mechanism and an incoherent squeezed thermal bath~\cite{Centrone2021}, as well as battery charging with local squeezing~\cite{Konar2022}, which point at its utility within quantum technological research.
\\

%===========================================================================
%===========================================================================
%===========================================================================
%===========================================================================
%===========================================================================
%===========================================================================
%===========================================================================
%===========================================================================

\noindent \textbf{Conclusion}\\
In conclusion, we have studied theoretically the prototypical bipartite form of a Gaussian quantum battery. We began by revisiting the case of a linearly driven battery charger, where we highlighted the crucial role of exceptional point physics in the ergotropic response of the battery. We then considered the arguably more interesting case of quadratic driving, where we found critical points instead play a decisive role in the battery energetics, including by influencing several unconventional features such as the spectral collapse of the Hamiltonian, quantum squeezing, dynamic instabilities and dissipative phase transitions. Our proposed quadratic quantum battery exhibits various desirable features, including storing only relatively small amounts of useless energy, allowing for the possibility for storing (theoretically unbounded) amounts of ergotropy, and requiring reasonable charging times to achieve significant energy storage. We hope that our theoretical proposal for a quadratic quantum battery can soon be realized with contemporary quantum platforms such as photonic cavities~\cite{Rota2019, Marty2021} and quantum circuits~\cite{Macklin2015, Nigg2017}, so that a squeezed battery may become a viable candidate for an energy storage device within the next generation of quantum technology. On the theory side, it should also be interesting to consider the scaling up of the quadratic quantum battery to include multiple battery cells~\cite{Campaioli2017, Gyhm2022} and cooperative effects~\cite{Martin2021, Jaseem2023}.
\\

%===========================================================================
%===========================================================================
%===========================================================================
%===========================================================================
%===========================================================================
%===========================================================================
%===========================================================================
%===========================================================================

\noindent \textbf{Acknowledgments}\\
\textit{Funding}: CAD is supported by the Royal Society via a University Research Fellowship (URF\slash R1\slash 201158) and by Royal Society Enhanced Research Expenses which support MSU. \textit{Discussions}: We thank V.~A.~Saroka for fruitful discussions.
\\

%===========================================================================
%===========================================================================
%===========================================================================
%===========================================================================
%===========================================================================
%===========================================================================
%===========================================================================
%===========================================================================

\noindent \textbf{Author contributions}\\
CAD conceived of the study and wrote the first version of the manuscript, with revisions from MSU. Both CAD and MSU performed the calculations and gave final approval for publication.
\\

%===========================================================================
%===========================================================================
%===========================================================================
%===========================================================================
%===========================================================================
%===========================================================================
%===========================================================================
%===========================================================================

\noindent \textbf{Competing interests}\\
The authors declare no competing interests.
\\

%===========================================================================
%===========================================================================
%===========================================================================
%===========================================================================
%===========================================================================
%===========================================================================
%===========================================================================
%===========================================================================

\noindent
\textbf{ORCID}\\
C. A. Downing: \href{https://orcid.org/0000-0002-0058-9746}{0000-0002-0058-9746}.
\\
M. S. Ukhtary: \href{https://orcid.org/0000-0001-5197-7354}{0000-0001-5197-7354}.
\\

%===========================================================================
%===========================================================================
%===========================================================================
%===========================================================================
%===========================================================================
%===========================================================================
%===========================================================================
%===========================================================================

%===========================================================================
%===========================================================================
%===========================================================================
%===========================================================================
%===========================================================================
%===========================================================================
%===========================================================================
%===========================================================================

\end{document}